\newif\ifPNAS
\PNASfalse
\PassOptionsToPackage{hyphens}{url}
\documentclass[9pt,twocolumn,twoside]{pnas-new}
\templatetype{pnasresearcharticle}
\ifPNAS\articletype{Anthropology}\fi
\usepackage{xcolor}
\newif\ifmarkchanges
\markchangesfalse

\usepackage{float}
\graphicspath{{Figures/}}
\usepackage{longtable}
\usepackage{enumitem}
\usepackage{multibib}
\usepackage{placeins}
\newcites{S}{SI References}
\DefineSpuriousJournalWord{The}
\DefineSpuriousJournalWord{for}
\DefineJournalException{Archiv f{\"u}r Orientforschung}{Arch. Orientforsch.}
\newif\ifwithSI
\withSItrue

\makeatletter
\newcommand{\journalname}{\ifPNAS\textbf{PNAS}\hspace{7pt}\fi}
\ifPNAS\else
  \patchcmd{\@maketitle}{\put(28.9,9.05){\color{typecol}\sffamily\bfseries\footnotesize RESEARCH ARTICLE}\put(102,3.55){\color{gray80}{\rule{0.5pt}{16pt}}}\put(110,9.05){\@articletype}\par\vskip11\p@}{}{}{\PackageError{PNAS_Paper}{The title page of pnas-new.cls changed}{}}
\fi
\fancypagestyle{firstnolines}{\fancyhf{}%
  \fancyfoot[L]{\footerfont\journalname\today}%
  \fancyfoot[R]{\footerfont\textbf{\thepage\ of \pageref*{LastPage}}}}
\@ifundefined{PNAS@linecount}{\newcount\PNAS@linecount\PNAS@linecount\@ne\relax}{}
\makeatother

\title{Automated, auditable dating of the Babylonian Astronomical Diaries with language models and astronomical tools}

\author[a,1,2]{Lucas Carrit Delgado Pinheiro}
\author[b,c,d,1,2]{Yuan-Sen Ting}
\author[e]{Jon Clindaniel}
\author[f]{Lu Li}
\author[a]{Yingbin Liang}
\author[a,g]{Ness Shroff}
\affil[a]{Department of Electrical and Computer Engineering, The Ohio State University, Columbus, OH 43210, USA}
\affil[b]{Department of Astronomy, The Ohio State University, Columbus, OH 43210, USA}
\affil[c]{Center for Cosmology and AstroParticle Physics (CCAPP), The Ohio State University, Columbus, OH 43210, USA}
\affil[d]{Max-Planck-Institut f\"ur Astronomie, K\"onigstuhl 17, D-69117 Heidelberg, Germany}
\affil[e]{Computational Social Science Program, Division of the Social Sciences, The University of Chicago, Chicago, IL 60637, USA}
\affil[f]{Centre for Astrophysics and Planetary Science, Racah Institute of Physics, The Hebrew University, Jerusalem, 91904, Israel}
\affil[g]{Department of Computer Science and Engineering, The Ohio State University, Columbus, OH 43210, USA}

\leadauthor{Carrit Delgado Pinheiro}

\ifPNAS\significancestatement{Historians and scientists rely on the Babylonian Astronomical Diaries, but many of their tablets have lost their date. Experts recover many such dates from the recorded sky and from other evidence. We built a system in which a rule-based parser and astronomical tools automate the comparison of the recorded sky with computed positions, while a language model checks the reading and adds what the parser missed. On 120 dated tablets, its leading candidate years include the published year in 96\% of cases. For each tablet, it shows how far the recorded astronomy alone narrows the year. The code and the evidence for each candidate year are open.}\fi

\authorcontributions{L.L. proposed the project and did early exploration. L.C.D.P. and Y.-S.T. designed research, performed research, built the system and wrote the paper. J.C. advised on archaeology, and Y.L. and N.S. advised on computer science.}
\authordeclaration{The authors declare no competing interest.}
\equalauthors{\textsuperscript{1}L.C.D.P. and Y.-S.T. contributed equally.}
\correspondingauthor{\textsuperscript{2}To whom correspondence may be addressed. E-mail: carritdelgadopinheiro.1@osu.edu or ting.74@osu.edu}


\begin{abstract}
The Babylonian Astronomical Diaries record the sky over nearly six centuries. Many tablets are broken and have lost their date formula. Experts dated many of them one tablet at a time in the standard edition, Astronomical Diaries and Related Texts from Babylonia (ADART). They used the recorded sky and other evidence, such as year formulas and royal names. Our system automates the astronomical part for all tablets with the same rules. A rule-based parser turns the recorded observations into clauses, short statements with fixed fields. A language model checks them against the text and adds observations that the parser missed. Astronomical tools test the clauses in all 701 years from 626 BCE to AD 75. The system ranks the years by how well they agree with the clauses. The system withholds published dates until its results are final, and the output lists the passages and computed positions behind each candidate year. On 120 dated tablets, the years tied at the top of the ranking include the published year in 96\% of cases. For 86 of them, this top group is a single year, the published year in 84. In 119 of the 120, the reported candidate set contains it. For 56 undated entries of ADART IV, the system gives ranked candidate years. For each tablet, the results show how far astronomy alone narrows the year. An interactive page at \url{https://tingyuansen.github.io/Babylonian_Dating/} follows each tablet and links the code, protocol and results.
\end{abstract}

\dates{This manuscript was compiled on \today}

\begin{document}

\maketitle
\thispagestyle{firstnolines}
\ifthenelse{\boolean{shortarticle}}{\ifthenelse{\boolean{singlecolumn}}{\abscontentformatted}{\abscontent}}{}

\Firstpage

\dropcap{S}cribes, mostly in Babylon, kept the Astronomical Diaries for nearly six centuries \cite{sachs1988adart1,sachs1989adart2,sachs1996adart3}. These cuneiform clay tablets report the Moon, the planets, eclipses and seasonal events, and many also report prices, river levels and political events. Historians of Persian, Seleucid and Parthian Babylonia use these reports \cite{pirngruber2013historical}, and economic historians use their prices, among the longest ancient price series \cite{temin2002price,pirngruber2017economy}. The observations also help to measure long-term changes in the Earth's rotation \cite{stephenson2016rotation}, and they include early records of aurora-like events \cite{hayakawa2016aurora}. Many of these uses depend on the date of each diary entry.

For many tablets, this date is hard to establish. Most tablets are now in the British Museum, and their original findspots are lost \cite{pirngruber2019museum}. Almost all tablets are also damaged, and the date formula is often lost \cite{sachs1988adart1}. When this formula survives, it normally fixes the year. When it is lost, the astronomical observations are the main evidence that remains.

Sachs and Hunger used this evidence to date many broken tablets in their edition, \textit{Astronomical Diaries and Related Texts from Babylonia} (ADART) \cite{sachs1988adart1,sachs1989adart2,sachs1996adart3,sachs2022adart4}. Volumes I to III hold the diaries with a published date, and Volume IV mostly the undated ones. They compared the surviving observations with computed planetary phenomena and eclipse catalogs, and they often combined several observations to separate candidate years. Difficult tablets needed further specialist calculation \cite{deJongHunger2020}. Such dating takes specialist time, and its reasoning is given in prose comments for each tablet. It is thus hard to see, for the corpus as a whole, how much of each published date rests on the astronomy. The reasoning is also hard to redo when a reading of the text changes, or when better calendars or astronomical tables appear.
\Endparasplit

Machine learning already automates parts of archaeology and philology, mostly by learning from examples \cite{Bellat2025MachineLearning,sommerschield2023survey}. It transliterates and translates Akkadian \cite{Gordin2020ReadingAkkadian,Gutherz2023TranslatingAkkadian}, classifies cuneiform tablets by period \cite{chen-etal-2024-classification}, and restores, dates and places Greek and Latin inscriptions \cite{assael2019restoring,assael2022restoring,assael2025contextualizing}. Agents built on language models can help build computational pipelines \cite{Swanson2025VirtualLab} and pass exact work to software tools \cite{Boiko2023Coscientist,Bran2024ChemCrow}.

Dating a diary needs such software tools, because planetary positions need exact computation. A language model may also have seen the diaries and their dates in training, and could repeat a date from memory. In our system, computed positions and fixed rules give the year, and a language model checks and completes the reading (Fig.~\ref{fig:workflow}).\footnote{A related project at the University of Chicago uses language models to extract the observations and a Bayesian model to check them against computed positions (\url{https://datascience.uchicago.edu/news/havi-award-supports-uchicago-researchers-pairing-ai-with-bayesian-statistics-to-decipher-the-babylonian-astronomical-diaries/}).}

In step 1, a rule-based parser reads the text into clauses, statements with fixed fields. The language model checks them against the text and adds the observations that the parser missed. Fixed rules treat broken text, restored text and unobserved events, and a written protocol records these rules for agents and human auditors (\textit{SI Appendix}, section~S2). In step 2, astronomical tools test the clauses in all candidate years. Agents can use the same tools through the Model Context Protocol (MCP), a standard for such calls \cite{mcp2026}. In step 3, fixed rules decide which years remain, and they give the same result on each run. The published dates enter after the candidate sets are final, and the output links each candidate year to its passages and computed positions.

For each tablet, we measure how far its surviving astronomy narrows the year. A planet seen in a zodiacal sign in a given month agrees with each year in which the planet is there in that month. Thus, one planet rarely fixes a year. A second planet with a different period acts as a second clock, and the years with which both clocks agree score higher. Lunar eclipses and the solar year act as additional clocks, and how far a tablet can be dated depends on which clocks survive. We make this measurement on 179 dated records from ADART I to III and on the undated entries of ADART IV.

\section*{Results}

\subsection*{Step 1. Reading the text into clauses}

\begin{figure*}[t]
\centering
\includegraphics[width=\linewidth]{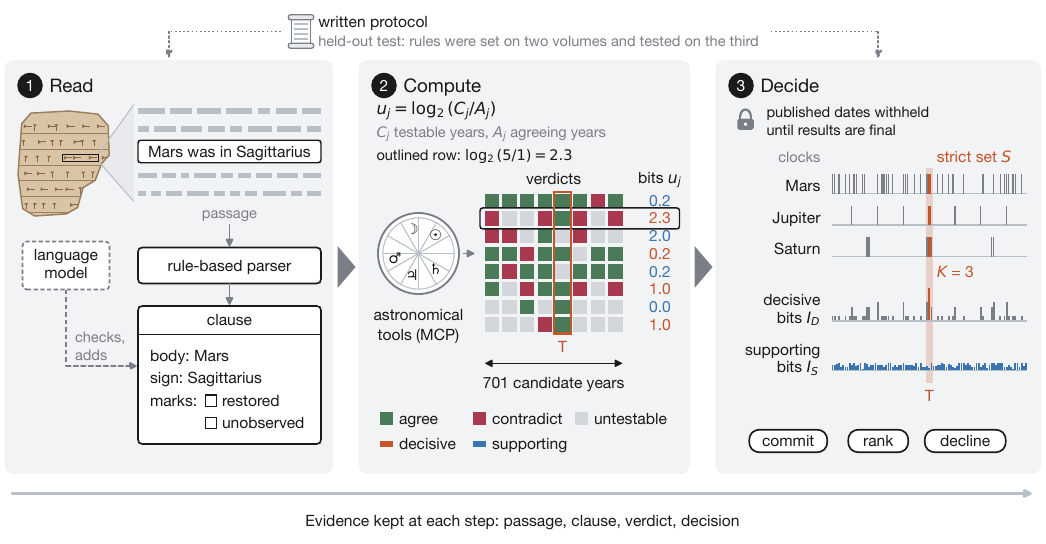}
\caption{The automated workflow, with a worked example. The banner at the top names the written protocol of the rules and the held-out test (\textit{SI Appendix}, sections~S2 and S8). In step 1, the parser turns a passage of the text into a clause, with its fields and marks. The language model checks it and adds the clauses that the parser missed. In step 2, astronomical tools, which agents can call through MCP, give each clause (grid row) a verdict in each candidate year. The grid shows eight clauses in eight of the 701 years. Bits $u_j$ follow from the verdicts of a row (Eq.~\ref{eq:weight}). Only decisive clauses (orange bits), whatever their bits, decide which years stay in the strict set. Supporting clauses (blue bits) only order the years. Column T is the one year with which all decisive rows agree. Step 3 shows 91 years of the example, and each clock row marks the years with which the clauses of that clock agree. Fixed rules count the decisive clocks $K$ that agree with each year and add up its decisive bits $I_D$ (Eq.~\ref{eq:info}). Year T leads on both and forms the strict set $S$ (Eq.~\ref{eq:strict}, orange). The supporting bits $I_S$ (blue) order the years. The system then commits to one year, ranks the years or declines (Eq.~\ref{eq:commit}).}
\label{fig:workflow}
\end{figure*}

\begin{figure*}[t]
\centering
\leavevmode\llap{\raisebox{0.264\textwidth}{\sffamily\bfseries\footnotesize a}\hspace{3pt}}%
\includegraphics[height=0.276\textwidth,trim=72 234 325.4 15.8,clip]{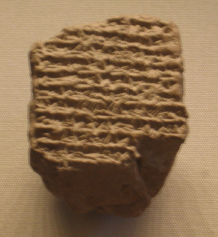}\hspace{6pt}%
\includegraphics[height=0.276\textwidth,trim=0 0 0 175,clip]{fig2_diary.pdf}
\caption{A diary tablet and the information in its clauses. (a) ADART I no.~$-322$B (British Museum, BM 45962), a diary for 323/322 BCE that records the death of Alexander the Great. Photograph by Shaunnol, Wikimedia Commons, public domain. (b) The opening of X301322 from ADART III, one tablet line per row, colored orange for a decisive and blue for a supporting clause. Square brackets mark text that Sachs and Hunger restored. Each log-scaled bar shows the number $n$ of years that agree with the row's strongest clause, labeled with $\log_2(701/n)$ bits (Eq.~\ref{eq:weight} gives the scoring). Gray rows hold unscored text, and their bars reach all 701 years (dashed line) at 0 bits.}
\label{fig:diary}
\end{figure*}

The system first turns each diary (Fig.~\ref{fig:diary}a) into clauses (Fig.~\ref{fig:workflow}). A clause is one astronomical report in fixed fields, such as the body, the phenomenon, the reference star or sign, and the month and day. For example, ``Mars was 1 finger above $\theta$ Ophiuchi'' gives one clause, and the parser passes over a weather report (Fig.~\ref{fig:diary}b). Each clause also records whether it rests on words that Sachs and Hunger restored. Another mark shows whether the scribe reports the event as unobserved (\textit{SI Appendix}, section~S2).

The rule-based parser is code with no model inside. It matches the common wordings of these reports and gives the same clauses on each run. In ADART I to III, it finds 3,193 clauses in 316 of the 417 records with a published date. A coding agent proposed some of its rules, for wordings that the parser often failed to read. Each accepted rule entered the code after tests and human review (\textit{SI Appendix}, section~S2).

A language model, GPT-6-Luna, then reviews each parser clause and searches the text for further observations. Candidate years and scores stay hidden from the model. Code also replaces the record identifier with an arbitrary label. It hides the year numbers and the names of kings in the translation (\textit{SI Appendix}, section~S2).

The model may keep, modify or drop each clause. A separate call must confirm each change by citing its passage, the exact words of the clause's own sentence that support it. Each added observation must cite a passage that code finds in the translation, and fixed checks then test all results against the clause format. The confirming call and these checks reject changes that the text does not support (\textit{SI Appendix}, Table~S1). A small model is enough, because the model reads and cites passages and code tests what it returns. In a comparison on ten records, the larger GPT-5.6-Terra cost an order of magnitude more. Its added clauses agreed with the published year about as often (\textit{SI Appendix}, section~S2).

The search adds clauses that the parser missed. It raises the 3,193 parser clauses to 4,010 clauses from 321 records (\textit{SI Appendix}, section~S2). Where the tools can test them, the reviewed clauses agree with the published years in 93.6\% of cases and the parser clauses in 93.8\%. The added clauses are thus about as reliable as the parser's (\textit{SI Appendix}, Table~S4). The dates in this paper therefore use the reviewed clauses. These are the parser clauses as the review leaves them and the clauses that the search adds.

\subsection*{Step 2. Checking clauses against the sky}

For each clause and each of the 701 candidate years, the tools return one verdict, agree, contradict or untestable (\textit{SI Appendix}, section~S4 and Table~S2). A verdict is untestable when the event went unobserved or when a clause misses a needed field, such as the day of a distance. A clause whose month is lost is tested over its whole year. The fewer years a clause agrees with, the more information it carries, which we measure in bits as
\begin{equation}
u_j=\log_2\!\left(\frac{C_j}{A_j}\right),
\label{eq:weight}
\end{equation}
where $C_j$ is the number of the 701 candidate years in which clause $j$ is testable and $A_j$ the number in which it agrees. For example, a clause that agrees with one testable year in four carries two bits, and one that agrees with all of them carries zero.

Each clause has two separate properties.
\begin{itemize}[leftmargin=*,itemsep=1pt,topsep=2pt]
\item Its bits, computed from its verdicts, say how much it narrows the years.

\item Its tier, decisive or supporting, says what it may do, and only decisive clauses decide which years remain. A decisive clause needs unrestored text and no mark of an unobserved event. It also needs a secure date or a test over its whole year, and a type that acts as a clock of the year. We chose the decisive types from a manual study of the clauses with help from language models. The choice used their agreement at the published dates of ADART I to III, their independence and their number of cases (\textit{SI Appendix}, section~S5). A supporting clause, such as a Moon position against a star, only orders the years.
\end{itemize}
The five clocks are Mars, Jupiter, Saturn, lunar eclipses visible at Babylon and the solar year. Eclipse reports are common, but most are untestable, because the scribe marks them as unobserved or their month and day are lost (\textit{SI Appendix}, section~S5). A clause of one clock agrees with a series of years that repeats with the clock's cycle. The solar year acts as a clock, because Babylonian months follow the Moon. The Babylonian date of a solstice, equinox or Sirius phase thus shifts from year to year. A recorded date agrees with the years that put the event on that day. Diaries take these dates from an arithmetical scheme, which the rules approximate (\textit{SI Appendix}, section~S4).

Clauses differ in information even within one tablet, as the opening of record X301322 shows, with its weather, lunar and planetary reports (Fig.~\ref{fig:diary}b). The first row states the year, which the system ignores while dating. Rows 2 and 4 place the Moon by a star on a stated night. On a given day of a Babylonian month, the Moon stands in about the same part of the sky each year. The Moon moves some $30^\circ$ in the test window of about two and a half days (\textit{SI Appendix}, section~S4). Rows 2 and 4 agree with 656 and 635 years. Mars passes a star about every two years, in a month that moves each time. The Mars row thus agrees with 11 years, which Fig.~\ref{fig:diary}b shows as 6.0 bits.

\begin{figure*}[t]
\centering
\includegraphics[width=\linewidth]{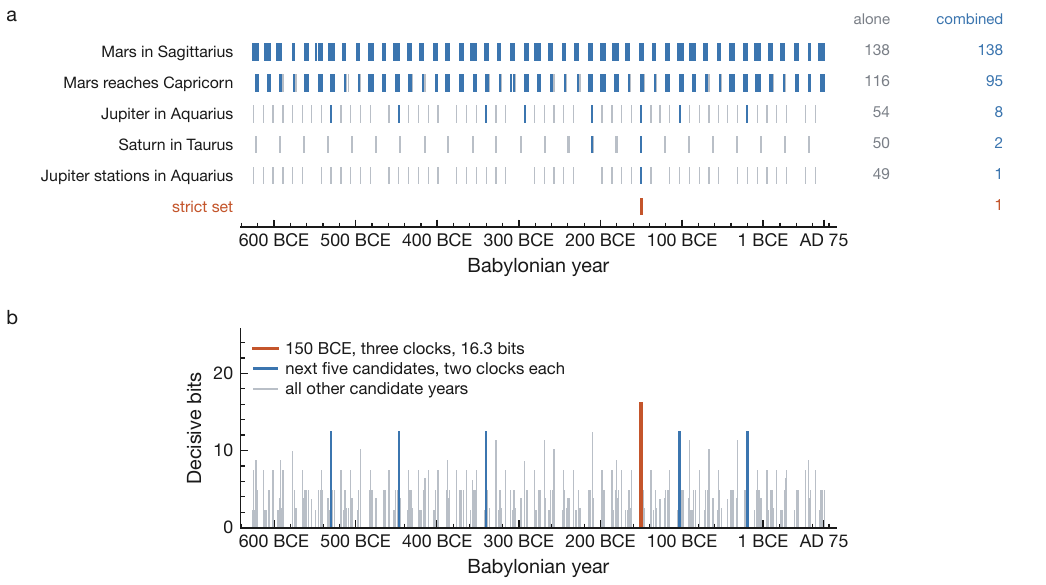}
\caption{Five decisive clauses of X301492 narrow the candidates to one year. (a) Gray ticks mark the years that each clause agrees with on its own. Blue ticks mark the running intersection, and the counts of both are at the right. The orange row is the strict set computed from $K$ and $I_D$, the single year 150 BCE. (b) Decisive bits $I_D$ of each candidate year. 150 BCE agrees with three clocks, and the five years with the next highest $I_D$ agree with two each.}
\label{fig:tablet}
\end{figure*}

The most common clause places the Moon or a planet against a Normal Star, a reference star near the path of the Sun. Such clauses agree with a median of 548 of the 701 years, or 0.4 bits. Eclipses and planets in a named sign carry several bits (\textit{SI Appendix}, Table~S3).

A clause with high bits can still be supporting. A station is where a planet appears to stop and turn back. A station against a star carries a median of 8.5 bits. However, it agrees with the published year in 48\% of testable cases, so it remains supporting (\textit{SI Appendix}, section~S5). Most of its contradictions fall outside the window of the clause. A planet looks stationary for several days, and the scribe's day may differ from the computed moment.

\subsection*{Step 3. Deciding with several clocks}

From the verdicts and tiers, the system decides in five stages (\textit{SI Appendix}, section~S6 and Table~S5).
\begin{enumerate}[leftmargin=*,itemsep=1pt,topsep=2pt]
\item It scores each year $y$ three ways. $K$ counts the decisive clocks that agree with the year. The bits of the agreeing decisive clauses add up to $I_D$, and those of the agreeing supporting clauses to $I_S$,
\begin{equation}
\begin{gathered}
K(y)=\sum_{c}\max_{j\in D_c} a_j(y),\\
I_D(y)=\sum_{j\in D} u_j\,a_j(y),\qquad I_S(y)=\sum_{j\notin D} u_j\,a_j(y).
\end{gathered}
\label{eq:info}
\end{equation}
Here $a_j(y)$ is 1 if clause $j$ agrees with year $y$ and 0 otherwise. $D$ holds the decisive clauses and $D_c$ those of clock $c$. In these sums, clauses with identical verdicts share their bits equally. The statement ``Saturn and Mars were in Leo'' gives one clause per planet, tested as one joint position. Its bits count once, and each planet counts for its own clock (\textit{SI Appendix}, section~S6).

\item It forms the strict set $S$, the years that no other year beats on the decisive evidence. Year $y'$ beats year $y$ if it has at least as many clocks and at least as many decisive bits, and strictly more of at least one,
\begin{equation}
\begin{aligned}
S=\{\,y:\ &\text{no } y' \text{ has } K(y')\ge K(y) \text{ and } I_D(y')\ge I_D(y),\\
&\text{with at least one inequality strict}\,\}.
\end{aligned}
\label{eq:strict}
\end{equation}
In the strict set, a contradiction counts as an untestable verdict, because about 6\% of testable clauses contradict even the published year (\textit{SI Appendix}, Table~S4). The number of contradictions enters only the ranking (Eq.~\ref{eq:rank}).

\item It forms the safety set. A source statement is the text behind one or more linked clauses, as for Saturn and Mars in Leo. Let $S^{(-s)}$ be the strict set recomputed without the decisive source statement $s$. The safety set joins the strict set and all these sets,
\begin{equation}
S_{\mathrm{safe}}=S\cup\bigcup_{s}S^{(-s)}.
\label{eq:safe}
\end{equation}
Here $\cup$ joins sets, so the safety set holds each year that one misread decisive statement could have pushed out of the strict set. Its other years rank right after the strict set.

\item It ranks each year $y$ by the ordered list of scores
\begin{equation}
r(y)=\bigl(t(y),\,K(y),\,I_D(y),\,I_S(y),\,-N_C(y)\bigr),
\label{eq:rank}
\end{equation}
where $t(y)$ is 2 in $S$, 1 in the rest of $S_{\mathrm{safe}}$ and 0 otherwise. $N_C(y)$ is the number of source statements that contradict $y$. Years are sorted by the first entry, then the second and so on, and the minus sign puts fewer contradictions first. Years with equal $r$ tie and form one group, and the groups $L_1,L_2,\dots$ run from the top of the ranking down.

\item It decides. If the strict set holds a single year $y^\ast$, so that $|S|=1$, the system commits to $y^\ast$ only if
\begin{equation}
E(y^\ast)\quad\text{or}\quad\bigl[\,K(y^\ast)\ge 2\ \text{and}\ |S_{\mathrm{safe}}|=1\,\bigr].
\label{eq:commit}
\end{equation}
Here $|\cdot|$ counts the years in a set, and $E(y^\ast)$ holds if an observed lunar eclipse agrees with $y^\ast$. A commitment thus needs an eclipse, or else two or more clocks and a safety set of one year. The system declines if the record has no decisive clause or $S$ keeps all 701 years, and it ranks the years otherwise.
\end{enumerate}

\begin{figure*}[t]
\centering
\includegraphics[width=\linewidth]{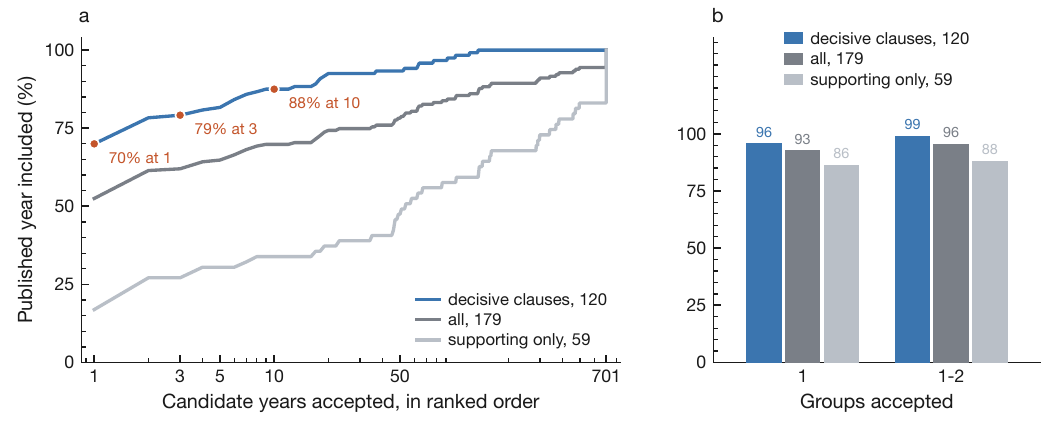}
\caption{Inclusion of the published year for the 120 benchmark records with decisive clauses, the 59 supporting-only records and all 179. (a) Percentage of records whose published year is among the first $n$ candidate years, on a logarithmic axis that ends at all 701 years. A record counts at $n$ only when each year ranked above or tied with its published year also lies within the first $n$. (b) Percentage of records whose published year is in the first one or two groups, each a set of tied years.}
\label{fig:performance}
\end{figure*}

\subsection*{Three clocks leave one year for record X301492}

Record X301492, one of 30 benchmark records with three or more clocks, shows how separate clocks narrow the candidates (Fig.~\ref{fig:tablet}). Five of its decisive clauses agree with 138, 116, 54, 50 and 49 years. They come from passages such as ``The 29th, Mars reached Capricorn''. Their running intersection falls to 95, 8, 2 and finally 1 year, 150 BCE, which is the published date (Fig.~\ref{fig:tablet}a).

The five clauses come from three planets, and so from three clocks, as Mars and Jupiter each give two clauses and Saturn one (Fig.~\ref{fig:tablet}a). Step 3 counts clocks apart from bits, because a second clause of the same planet adds bits to the same clock. At 150 BCE all three clocks agree, and the decisive bits $I_D$ total 16.3 (Fig.~\ref{fig:tablet}b). The five years with the next highest $I_D$ agree with two clocks each. The strict set thus holds 150 BCE alone, as does the intersection.

Even so, step 3 ranks 150 BCE first but does not commit to it. A commitment on clocks alone needs a safety set of one year (Eq.~\ref{eq:commit}). For this record, removing one decisive statement at a time adds six other years, and the safety set holds seven.

\subsection*{Tests on a benchmark of 179 dated records}

To test the system on many tablets, we fixed a benchmark of 179 dated records. It holds records that had at least three parser clauses when we fixed it, before comparing any candidate set with a published date. Of these, 120 have at least one decisive clause, and we call them records with decisive clauses. The other 59 are supporting-only records (\textit{SI Appendix}, section~S1). Step 3 declines supporting-only records, but the benchmark still scores their ranking. We score each reported set by its inclusion rate, the share of records whose published year lies in the set (Fig.~\ref{fig:performance}).

The first group, $L_1$, holds the top-ranked years. For the 120 records with decisive clauses, it contains the published year in 115 (96\%, Fig.~\ref{fig:performance}b). For 86 of the 120, it holds a single year, and in 84 this is the published year. The first two groups contain it in 119 (99\%), and the third and fourth add none. The supporting-only records also reach 86\% for the first group, but their groups hold many years (Fig.~\ref{fig:performance}a).

In Fig.~\ref{fig:performance}a, the first candidate year alone is the published year in 70\% of the records with decisive clauses. Ties explain the gap to the 96\% of the first group. The published year can tie with other years on all five scores of Eq.~\ref{eq:rank}.

The strict set $S$ (Eq.~\ref{eq:strict}) is the reported set of candidate years. It contains the published year in 119 of the 120 records with decisive clauses (99\%), with a median size of 15 years. In the one miss, X301582, the published year stays in the safety set.

The published dates set only a few general choices, such as the decisive types, some parser rules and the tolerances. The room for overfitting is thus small. A held-out test chooses these again on two volumes and dates the third, so each volume is the test volume once (\textit{SI Appendix}, section~S8). Over the three rounds, the first group holds the published year in 96\% and the strict set in 99\%, as in the full benchmark. On the benchmark, most of the 31 alternative settings of the method change the rates little. Removing the tiers lowers inclusion in the strict set (\textit{SI Appendix}, section~S7).

A registered test on dated diaries new to the system gives similar rates. ADART IV adds such diaries, the addenda, and we fixed the rules and the scores of the test before reading them. Of the 39 addenda that do not join (fit onto) or duplicate a record of ADART I to III, 12 have a decisive clause. For 11 of them, the first group and the strict set each hold the published year (\textit{SI Appendix}, section~S8).

The same rules date the 238 dated records outside the benchmark, and 29 of them have a decisive clause. Their strict sets hold a median of 87 years, against 15 in the benchmark, and 28 of them hold the published year.

In the whole corpus of 417 dated records, the commitment rule of Eq.~\ref{eq:commit} commits to one year for 10. All 10 match the published year, and all 10 lie in the benchmark. Nine rest on two or more clocks and one on a lunar eclipse. A year that the tablet itself states gives a second check. After dating, the system finds a year stated in the Seleucid Era for 23 dated records, alone or with its Arsacid equivalent (\textit{SI Appendix}, section~S1). In all 14 records where the astronomy narrows the year, the stated year lies in the strict set.

\subsection*{Ranked candidate years for 56 undated entries}

With the rules tested on dated records, we applied them to the undated diaries published by ADART IV \cite{sachs2022adart4}. The system received only the translated astronomical text of each entry. The parser and the search find astronomical clauses in 399 entries, and 56 of them have a decisive clause and receive a ranked set. Their strict sets cut the 701 candidate years to a median of 102 (\textit{SI Appendix}, section~S9 and Table~S7). Most of these entries keep a single decisive clause that stands before any month heading, so its test covers a whole year.

Many of these strict sets repeat at a regular spacing (\textit{SI Appendix}, Fig.~S2), as step 2 predicts for a record with one clock. A clause from a second clock, for example on a fragment joined to the tablet, could narrow these series.

\section*{Discussion and Conclusion}

The system narrows the date of each diary from its astronomical records. It commits to a single year for 10 of the 417 dated records. For the others, it reports ranked candidate years with the evidence behind each. Sachs and Hunger also used surviving year formulas, royal names and the form of the script \cite{sachs1988adart1}. The output therefore shows, for each tablet, how far the astronomy alone narrows the year and where other evidence is necessary.

Specialists compared each diary's recorded sky with computed positions tablet by tablet, and we have automated this comparison. The same rules apply to all tablets, so a new reading, join or calendar can be tested on the whole corpus in one rerun. The review, the search and the dating of the 179 benchmark records cost about \$3. With the evidence open, specialists can test the candidate years of undated ADART IV entries against joins and historical reports. Other sources that record the sky could be dated in the same way, such as Greek horoscopes \cite{neugebauer1959horoscopes} and eclipse reports in Chinese chronicles \cite{stephenson2016rotation}. The same division of labor could help other archaeological work, such as calibrating published radiocarbon dates \cite{reimer2020intcal20}.

\matmethods{
\subsection*{Corpus}
We study the Diaries themselves, all of which ADART I to IV publish \cite{steele2024newmoon}. Astronomical Diaries Digital (ADsD), an open digital edition, holds ADART I to III and the other genres of ADART V and VI \cite{adsd}. We use its 417 records of ADART I to III with a published date. Of these, 393 have edition text and 332 an English translation. We read ADART IV, which is outside ADsD, from a scan of the printed volume with optical character recognition (\textit{SI Appendix}, section~S9).

\subsection*{Step 1. Reading the text}
The parser takes the month and day from headings in the text, such as ``Month VI'' or ``Night of the 7th''. The language model of the review and search is gpt-6-luna-2026-09-22, called with low reasoning effort and a fixed output schema (\textit{SI Appendix}, section~S2). Earlier versions of the system, from 2024, used GPT-4o and then GPT-5 in the same role.

\subsection*{Step 2. Calendar, ephemeris and rules}
The candidate calendar holds the 701 Babylonian years from 626 BCE to AD 75, with the month starts of Parker and Dubberstein \cite{parker1956babylonian}. We label each Babylonian year by the Julian year in which it begins, so the year 150 BCE runs into 149 BCE. The tables of Parker and Dubberstein begin in 626 BCE. The calendar thus leaves out 652 to 627 BCE, the years of the earliest diaries and of one dated record. Daily positions of the Sun, the Moon and the five visible planets come from the Development Ephemeris DE441 of the Jet Propulsion Laboratory (JPL) \cite{park2021jpl}. This is a standard table of computed positions. We read it with jplephem and compute eclipse circumstances at Babylon with the Swiss Ephemeris (\textit{SI Appendix}, section~S4). The Babylonian zodiac is fixed to the stars, so we convert a longitude of date to this frame with the relation of Britton \cite{britton2010zodiac},
\begin{equation}
\lambda_{\mathrm{Bab}}=\lambda_{\mathrm{date}}+3.20^\circ-1.3828^\circ\,Y,
\label{eq:britton}
\end{equation}
where $Y$ is the number of Julian centuries from astronomical year 0. Each sign then covers $30^\circ$ of $\lambda_{\mathrm{Bab}}$, from Aries at $0^\circ$.

Most rules test a clause in the Julian interval of its Babylonian month (\textit{SI Appendix}, sections~S3 and S4 and Table~S2). Distances use one cubit $=2^\circ$ and 24 fingers $=1$ cubit, with tolerances of $3^\circ$ for the Moon and $2^\circ$ for a planet. A lunar eclipse must match a catalog eclipse that Babylon could see \cite{espenak2009lunar}. Its visibility depends on $\Delta T$, a correction of three to five hours for the slowing rotation of the Earth. The catalog gives $\Delta T$ for each eclipse. The eclipse must be visible with this value and with one standard error either side (\textit{SI Appendix}, section~S4).

\subsection*{Blinding}
Published dates, volume, joins, catalog order and other fields that carry date information are withheld until the clauses, verdicts and candidate sets are final. While dating, the system uses year statements inside a tablet only for the difference between two year headings (\textit{SI Appendix}, section~S1).\par
}

\showmatmethods{}

\acknow{L.L. raised the question of this project at the Hebrew University, in discussions with an archaeologist friend there. L.L. brought it to Y.-S.T., who assembled the team. Coding agents were then weak, and Y.-S.T. used the project as a test problem for them. By 2025, agents could do part of the work, but their output needed much human checking. L.C.D.P., who led a national team at the International Olympiad on Astronomy and Astrophysics (IOAA), then joined to build and check the system. IOAA problems test exactly this kind of positional astronomy. Galactic archaeology, the research field of Y.-S.T., uses stars to study the history of the Milky Way. J.C., a dormitory mate of Y.-S.T. in graduate school, joined after more than a decade of debate over whether it counts as archaeology. The work is independent of the Chicago project noted in the introduction. The manual checks of the clauses and rules took more than a year. Y.-S.T. is supported by a Humboldt Fellowship of the Alexander von Humboldt Foundation. Claude Opus 4.6 to 5.5 (Anthropic) helped develop the code and helped draft and edit the text. The authors reviewed all of the text and code, and L.C.D.P. verified the code.}

\showacknow{}

\dataavail{The interactive page at \url{https://tingyuansen.github.io/Babylonian_Dating/} follows each record and links the repository, \url{https://github.com/tingyuansen/Babylonian_Dating}. The repository holds the code, the protocol, the clauses, the benchmark, the results of all records and the tables read from Parker and Dubberstein.}

\bibliography{Bibliography}

\ifwithSI
\clearpage
\onecolumn
\raggedbottom
\renewcommand{\topfraction}{0.85}\renewcommand{\textfraction}{0.1}\renewcommand{\floatpagefraction}{0.85}
\setcounter{figure}{0}\renewcommand{\thefigure}{S\arabic{figure}}
\setcounter{table}{0}\renewcommand{\thetable}{S\arabic{table}}
\setcounter{equation}{0}\renewcommand{\theequation}{S\arabic{equation}}\renewcommand{\theHequation}{S\arabic{equation}}
\setlist{leftmargin=*,itemsep=1pt,topsep=3pt}

\begin{center}
{\sffamily\bfseries\Large SI Appendix}\\[4pt]
{\sffamily Automated, auditable dating of the Babylonian Astronomical Diaries with language models and astronomical tools}
\end{center}

\subsection*{S1. Corpus, input and benchmark}

The scores of the main text rest on a fixed set of records and published years, which this section defines. We use the ADsD files, distributed under a CC0 license through the Open Richly Annotated Cuneiform Corpus (ORACC). A record with edition text holds a transliteration and, where available, an English translation.

The input of each record holds its ADsD number and the English text. The ADsD number encodes the published year, as X103222 does for ADART I no.~$-322$B. The dating uses it as the name of the record, and the language model sees an arbitrary label in its place (section~S2). Year statements also appear in the text, such as ``Year 115, which is year 179, king Arsaces'' in X301322. Apart from the difference between two year headings (section~S2), the system reads them only after dating. The check after dating finds a Seleucid Era year, alone or with an Arsacid Era year, and converts it as Parker and Dubberstein do \cite{parker1956babylonian}.

The benchmark holds the records that had at least three parser clauses when we fixed it. A record has decisive clauses when its result keeps at least one. Audits of disagreements between the astronomy and the published dates led to changes in some parser rules, context rules and the calendar transcription. They left the published dates and the list of benchmark records as they were.

The published year of a record is the year of its ADsD date. A tablet that covers two years is scored by its first year, because each candidate year is the year in which a tablet begins. The clauses after a year change are tested in the following year (section~S2).

ADART IV makes 27 corrections to Volumes I to III. The three that change a published date are already in ADsD, and the others leave the scored clauses unchanged. The benchmark thus holds 179 records with fixed published years.

\subsection*{S2. Reading the text (step 1)}

Every date rests on clauses, so this section gives the rules that read them and the checks on the language model. Besides the restored and unobserved marks, a clause holds the phenomenon, the bodies, the star or sign, and the measurement with its unit and direction. It also holds the month, day and year, and the position of its passage in the translation. The parser rules, with the wordings that they match and the names of stars and signs, are part of the released code. The date fields come from headings in the text, by these rules.
\begin{itemize}
\item A month heading such as ``Month VI'' gives its month to the reports that follow, until the next month heading.

\item A report usually runs through the days of one month. Sometimes the day number drops by more than ten, for example from the 28th to the 3rd, without a new month heading. A later month has then begun, but its heading is lost, and more than one month may have passed. The position clauses after the drop thus keep no month. They are tested over their whole year and stay supporting (section~S5).

\item A day heading such as ``Night of the 24th'' gives its day to the measurements after it. A measurement takes that day only if the heading stands shortly before it, in intact text of the same sentence. Otherwise its day is uncertain and stays empty. Eclipse, solstice, equinox, Sirius and acronychal clauses take their day from the last day heading by a similar rule.

\item In the edition's translation, square brackets enclose words that Sachs and Hunger restored where the tablet is broken, as in ``solar [eclipse'' (Table~\ref{tab:llm}). A clause counts as restored if its body, sign, star, number, unit or direction lies in brackets. It also counts as restored if the parser can read it only across a bracket. A clause keeps its heading across brackets.

\item Many reports end a month with a summary, which often begins with ``That month''. A position clause in such a summary takes the month that ends at the next month heading. The same rule gives a month to a clause just before the first heading of a tablet.

\item Some tablets run from the end of one year into the next. A year heading marks the change, and so does a month number lower than the one before. Clauses after the change are tested in the year after each candidate year. A later year heading moves the clauses after it by the difference between its number and that of the first year heading.
\end{itemize}

The model works in two passes, a review of each parser clause and a search for missed observations. In the review, it receives one parser clause with 500 characters of the translation on each side. Each request holds an arbitrary label in place of the ADsD number, and placeholders replace the year numbers and the names of kings. Fixed code checks each proposal, and the prompts and model settings are archived with the code.
\begin{itemize}
\item A modification may change the phenomenon, a body, the star or sign, or the measurement. It may also change the direction or the link to the body named. When the clause's sentence holds a bracket of Sachs and Hunger, it may propose the restored mark, and a field check confirms the bracket. Each modification is split into changes of single fields, which are confirmed and accepted one by one.

\item A clause takes its month, day and year from the headings before it. The model may replace a month or a day with a value that the clause's own sentence states. A month, a day or a restored mark that the headings give always stays, and code sets the year.

\item Scribes mark some events as not observed, with phrases such as ``I did not watch'' or ``I did not see''. Code marks such eclipses, stations and acronychal risings as unobserved, which makes their verdict untestable, and only code sets this mark. Solstice, equinox and Sirius dates come from the scheme, and they stay testable.

\item The parser splits some statements into linked clauses on purpose, and the model cannot drop one of these. To drop any other clause, a separate model call must cite a passage that gives a different value for one of its fields.

\item The search lists each missed observation with its passage, and a second call writes it in the clause format. Code must find this passage word for word in the translation. The passage must also name the body that the clause describes, or refer back to one named shortly before. Code also checks that the body, star, sign, measurement and direction fit each other, and that the type is one of Table~\ref{tab:rules}. An added clause counts as restored if its passage holds a bracket or the editors' question mark. The parser, too, reads such words only as restored or not at all.

\item Malformed output, or a passage absent from the translation, leaves the clause unchanged.
\end{itemize}

The clause counts of the main text include the linked clauses of each statement, so they exceed the number of separate observations. In the review, the model kept 3,142 parser clauses and modified the other 51. In the search, 795 of its 8,123 proposed observations passed the checks, and they gave 817 clauses. The 3,193 reviewed parser clauses and these 817 clauses make the 4,010 clauses of the main text. For ADART IV, the model kept 533 of the 539 parser clauses and modified the other 6, and the search added 179 clauses. Most of the other proposals state astronomy outside the rules, such as a planet's invisibility, or repeat a parser clause. Table~\ref{tab:llm} gives examples of each outcome.

A comparison on ten records also ran GPT-5.6-Luna and the larger GPT-5.6-Terra with the same prompts and checks. The added clauses of both agree at the published year in 89\% of testable cases, against 88\% for GPT-6-Luna. GPT-5.6-Terra cost about 20 times as much as GPT-6-Luna, so the 179 benchmark records would have cost about \$60 with it, against about \$3.

A coding agent, Claude Opus 5 (Anthropic), proposed some of the parser's reading rules. It did not review clauses, which is the task of GPT-6-Luna. It saw wordings that the parser often failed to read, with their records, and proposed a reading rule for each. Record identifiers and year statements were removed. Names of kings stayed in the text. Candidate years and scores stayed hidden from the agent.

After human review, the accepted proposals entered the code as general rules, and the authors wrote the code of each. Each rule had to pass focused tests and a check of its clauses at the published dates of ADART I to III (section~S5). Without the coding agent's rules, each rate falls by at most two points, and the median strict set grows by about a third (Fig.~\ref{fig:ablation}). The rules thus add reading coverage, and the rates do not rest on them.

A written protocol, free of dates and results, states the rules of steps 1 to 3. It is archived with the code and given unchanged to each agent and human auditor.

\begin{table}[H]
\centering
\caption{Examples from the review and search steps.}
\label{tab:llm}
\small
\renewcommand{\arraystretch}{1.15}
\begin{tabular}{p{0.17\linewidth}p{0.30\linewidth}p{0.45\linewidth}}
\toprule
Outcome & Record and passage & Result \\
\midrule
Parser clause modified & X300993, ``Around the 6th, Mars' last appearance in Gemini'' & The parser read the appearance without its day. Review proposed day 6 from the same sentence, and the field check accepted it. \\
\addlinespace
Restored mark added & X202270, ``solar [eclipse'' & The parser did not mark the clause as restored. Review cited the bracket before ``eclipse'', and the field check accepted the mark. \\
\addlinespace
Proposed change rejected & X202552, ``moon was] 1 1/2 cubits behind $\eta$ Tauri'' & Review proposed month V from a heading in an earlier sentence. The confirming call rejected it, because the month came from another sentence. \\
\addlinespace
Missed observation recovered & X301323, ``the moon having passed a little to the east, it stood 3 cubits in front of Venus to the west'' & The parser missed this relation, whose subject is ``it''. The search found it, and code bound ``it'' to the Moon named earlier in the sentence. \\
\addlinespace
Astronomical but unsupported & X102732, ``Mercury, which had set, was not visible'' & The search found an astronomical report, but the invisibility of a planet lies outside the rules. It was recorded in a separate category. \\
\addlinespace
Search proposal rejected & X102812, ``[\ldots{} Ni] ght of the 21st'' & The search proposed the Moon in front of $\alpha$ Virginis and cited this heading with a wrong letter. Code did not find the citation in the translation and rejected the proposal before any astronomical test. \\
\bottomrule
\end{tabular}
\end{table}

\subsection*{S3. Calendar (step 2)}

Every test needs the Julian dates of a Babylonian month in each candidate year, which this section takes from Parker and Dubberstein. Each Babylonian year is labeled by the Julian year in which it begins, so the last year, AD 75, runs into AD 76. Babylonian dates are converted to the Julian calendar, extended back before its introduction. A Babylonian day began at sunset, so it maps to the Julian date of its daylight part. Years are numbered astronomically, with 1 BCE as year 0.

The calendar stores the month starts, the number of months in each year and the type of intercalary month. An intercalary month is an extra month that keeps the calendar in step with the seasons. A second Ululu (VI$_2$) follows month VI, so months VII to XII of that year move one month later. A second Addaru (XII$_2$) follows month XII and leaves the earlier months of the same year in place. In a candidate year without that extra month, a clause dated in it is untestable.

We transcribed the tables of Parker and Dubberstein \cite{parker1956babylonian} with optical character recognition. Checks against the computed new moons and the printed table found 10 transcription errors, which we corrected. The month starts then fall a median of 1.7 days after the computed new moons, as months that begin at first crescent should.

\subsection*{S4. Astronomical rules (step 2)}

Every verdict comes from a test of one clause in one candidate year, and this section states these tests. Geocentric positions come from a grid computed from the JPL Development Ephemeris DE441 at Julian noon each day. Cubic splines interpolate between days. Positions on the grid are in barycentric dynamical time (TDB), a uniform time scale. The baseline uses these positions without the $\Delta T$ correction of the eclipses, because the windows of the rules span a day or more (section~S7). Normal Star positions come from the table in the introduction of ADART I, reduced to J2000.

The tools are written in Python. They read DE441 with the jplephem package, and SciPy gives the splines and the times of stations and sign entries. The Swiss Ephemeris, through pyswisseph, gives the local circumstances of eclipses. Precession to the ecliptic of date follows the formulas of Lieske as given by Meeus. The MCP server uses the MCP Python SDK, and the repository lists all versions.

The tools use two frames of ecliptic longitude, because the signs are fixed to the stars while the seasons are not. Distances to Normal Stars and between bodies use the fixed ecliptic of the year 2000 (J2000). The sign tests use the longitude of date, converted with Eq.~\ref{eq:britton} of the main text. Solstices and equinoxes use the frame of the mean equinox of date. Each is the moment when the Sun's longitude in this frame reaches $0^\circ$, $90^\circ$, $180^\circ$ or $270^\circ$.

Table~\ref{tab:rules} lists the test for each clause type. The tools compute stations and sign entries directly from DE441.

The eclipse, solstice, equinox, Sirius and acronychal rules test a stated day. Measured relations without a day are untestable, except a station against a star, which is tested over its month. The windows of 4 days, 3 days and $10^\circ$ are half widths.

A planet in a sign must be there at some time in the month. The month runs from the civil day before its first day, which holds the opening sunset, to the start of the next month. A clause whose month is lost is tested over its whole year. A station must lie in its stated sign.

The diaries take their solstice, equinox and Sirius dates from a Babylonian computing scheme. A report such as ``The 19th, equinox'' in X301322 gives such a computed date. Solstice and equinox dates run about two days late, the median offset at the published dates. Their rule compares the stated date with the computed date shifted by two days.

The Sirius dates add the fixed intervals of the scheme \citeS{sachs1952sirius} to the shifted summer solstice, counted in months of 30 days. They are 21 days for the first appearance and 6 months and 11 days for the acronychal rising. The last appearance follows after 10 months and 27 days, in the next year.

\begin{table}[H]
\centering
\caption{Astronomical rules. Each rule returns agree, contradict or untestable for one clause in one candidate year. Here the month is the Julian interval of the Babylonian month that applies to the clause. Examples are passages of the ADsD translation.}
\label{tab:rules}
\small
\renewcommand{\arraystretch}{1.15}
\begin{tabular}{@{}p{0.18\linewidth}p{0.32\linewidth}p{0.13\linewidth}p{0.29\linewidth}@{}}
\toprule
Clause type & Test & Window or tolerance & Example \\
\midrule
Planet in a named sign & The planet is in the stated Babylonian sign during the month & The month & ``Mars was in Sagittarius'' (X301492) \\
First or last appearance in a sign & The planet is in the stated sign. The date of visibility is not tested & The month & ``Mercury's last appearance in the east in Libra'' (X301492) \\
Planet station & The planet's apparent motion along the ecliptic reverses & The month $\pm$ 20 days & ``Jupiter became stationary in Aquarius'' (X301492) \\
Station against a Normal Star & A station of the stated kind lies at the measured distance from the star & The month, or $\pm$2 days of a stated day, and $2^\circ$ & ``when Saturn became stationary to the west, it became stationary 1 cubit behind $\beta$ Geminorum'' (X300873) \\
Planet enters a sign & The planet crosses into the stated sign, in either direction of motion & The month $\pm$ 20 days & ``Mars reached Capricorn'' (X301492) \\
Moon against a star or planet & The relation holds in ecliptic longitude & $3^\circ$ & ``the moon was 2 1/2 [cubits] below $\alpha$ Librae'' (X301322) \\
Planet against a star or planet & The relation holds in ecliptic longitude & $2^\circ$ & ``Mars was 1 finger above $\theta$ Ophiuchi'' (X301322) \\
Eclipse of a stated type & A catalog eclipse is visible at Babylon in all three $\Delta T$ trials & $\pm$1.5 days & ``Month VII, night of the 14th, lunar eclipse in Aries'' (X300993) \\
Solstice or equinox on a day & The scheme date \citeS{neugebauer1948solstices} matches the solar longitude of date, after the two-day lateness of the scheme & 4 days  & ``The 19th, equinox'' (X301322) \\
Sirius date on a day & The scheme date \citeS{sachs1952sirius} matches the same solar computation & 3 days  & ``The 29th, Sirius' first appearance'' (X202250) \\
Acronychal rising on a day, the rising of a planet at sunset & The planet is opposite the Sun & $10^\circ$  & ``Saturn's acronychal rising'' (X301492) \\
\bottomrule
\end{tabular}
\end{table}

A statement such as ``Saturn and Mars were in Leo'' says that both planets were in the sign at the same time. Tested apart, the two clauses would also accept years in which the planets were there in different months. The tools therefore test them as one joint position. They search the day of the clause, or its month, for a stretch of time when both planets are in the sign. A joint position needs its month, while a single position can fall back on its whole year. A joint position with Mercury or Venus stays supporting.

A cubit has 24 fingers \cite{sachs1988adart1}, and the rules take it as $2^\circ$. Distances measured in the diaries give about $2.2^\circ$ for a cubit \citeS{fatoohi1998angular}, and section~S7 tests this value. A measured relation agrees in a candidate year if $|\Delta-d|\le\tau$ on the date of the clause. Here $d$ is the stated distance in degrees, $\Delta$ the computed longitude difference in the stated direction, and $\tau$ the tolerance.

The lunar tolerance is wider, because the nearby Moon appears shifted from Babylon, an effect called parallax. Each value is the smallest multiple of $0.5^\circ$ that covers 89\% of the differences between stated and computed distances of its body class. These differences come from the published dates of ADART I to III, which include the benchmark, so section~S7 also tests other values.

``In front of'' and ``behind'' are measured in ecliptic longitude. ``Above'' and ``below'' state a latitude. For them, the body's longitude must lie within the tolerance of the star's longitude. Phrases such as ``low to the south'' lack a distance and stay untested. A measurement on a stated day is tested from the midnight that begins the civil day before it to noon of the day after it. This window holds the night that opens the Babylonian day and one day of calendar uncertainty.

The eclipse catalogs are those of Espenak and Meeus for lunar \cite{espenak2009lunar} and solar eclipses \citeS{espenak2009solar}. The window of $\pm$1.5 days around the Julian date of the Babylonian day allows for the sunset start of the day and for calendar uncertainty. The catalogs give $\Delta T$ for each eclipse. It is uncertain for antiquity, so the rule tests this value and one standard error either side, 4 to 8 minutes for the diaries. Visibility is computed for Babylon, at $32^\circ32'30''$ N and $44^\circ25'30''$ E \citeS{pedersen2021babylon}. A body is visible if its center stands above an apparent altitude of $0^\circ$ at Babylon.

An eclipse of either kind agrees if it is of the stated type and visible at Babylon for all three values of $\Delta T$. A clause records the type of an eclipse, lunar or solar. The size of the eclipse, total or partial, stays outside the test. For a lunar eclipse, part of its umbral phase must be visible. If the window holds no catalog eclipse of the stated type, the clause contradicts the year. It also contradicts the year if an eclipse occurs but no phase is visible at Babylon in any of the three trials.

An eclipse whose visibility changes between the values is untestable. So is a lunar eclipse of which only the penumbral phase, a faint dimming in the Earth's partial shadow, is visible at Babylon. Solar eclipses stay supporting, because the dated corpus has only one unrestored, observed report of them with month and day.

\subsection*{S5. Evidence tiers and agreement at the published years (step 3)}

A misread decisive clause could remove the true year from the strict set. The decisive tier therefore takes the types that seldom contradict the published years. A supporting clause changes the order of the years, and it may be of any type that the tools can test.

The decisive types are these.
\begin{itemize}[leftmargin=*,itemsep=1pt,topsep=2pt]
\item Mars, Jupiter and Saturn, in each phenomenon that the tools test, except last-appearance clauses, stations against a star and distances to another planet.

\item Lunar eclipses.

\item Solstice, equinox and Sirius dates, which together form one clock, the solar year.

\item Distances between the Moon and Mars, Jupiter or Saturn, such as ``the moon was 2 1/2 [cubits] below Saturn'' in X301322. Such a distance counts for the clock of its planet. Its body, number, unit and direction must be unrestored. The example is supporting in Fig.~\ref{fig:diary}b, because its unit is restored.
\end{itemize}
A distance may be in one unit or in cubits and fingers. All other types are supporting. These include solar eclipses, the Moon against a star, Venus and Mercury, and types with too few testable examples at the published dates. Venus and Mercury stay supporting, because they stay close to the Sun and their position in a given month repeats on short cycles. A last appearance depends on visibility near the Sun, and a model of visibility based on elongation alone failed its held-out check. Last-appearance clauses stay supporting too.

The date of a decisive clause must be secure, unless its rule tests the whole year. A date is secure if it has a month, from a heading, the summary rule or a year change (section~S2). A rule that tests a day, such as an eclipse or a measured distance, needs a day heading for the clause. Its day number and month numeral must be unrestored. A position in a sign is tested over its whole month instead, and it stays decisive when only its month numeral is restored.

A position after a month heading lost at a drop in the day number (section~S2) is supporting. A position of Mars, Jupiter or Saturn before any month heading stays decisive, and its rule tests the whole year. All 27 such positions in ADART I to III that are testable at the published year agree with it.

Scribes mark some eclipses as omitted, excluded, passed by, unwatched or unobserved, and such a report is untestable. Of 95 eclipse clauses, the scribes mark 51 as unobserved, and 39 lack a secure month and day. An eclipse whose clause, month or day is restored counts only as supporting.

\begin{table}[H]
\centering
\caption{Information and decisive share by observation type, for 4,010 clauses from 321 records of ADART I to III. Clauses counts all clauses of the type, and Testable those testable in at least one of the 701 years. Years and Decisive count the testable clauses, and each row has at least five of them. Years is the median number of candidate years with which such a clause agrees. Bits are $\log_2(701/m)$ for the median $m$, while the scoring uses Eq.~\ref{eq:weight} of the main text. Decisive is the percentage of the testable clauses that meet the conditions of the decisive tier.}
\label{tab:information}
\small
\setlength{\tabcolsep}{3pt}
\begin{tabular}{@{}p{3.55cm}rrrrr@{}}
\toprule
Observation type & Clauses & Testable & Years & Bits & Decisive (\%) \\
\midrule
Moon or planet against a Normal Star & 2062 & 1330 & 548 & 0.4 & 1 \\
Planet in a named sign & 683 & 659 & 123 & 2.5 & 35 \\
Moon against a planet & 357 & 219 & 84 & 3.1 & 19 \\
First appearance in a sign & 229 & 229 & 436 & 0.7 & 15 \\
Last appearance in a sign & 222 & 222 & 534 & 0.4 & 0 \\
Planet enters a sign & 101 & 101 & 460 & 0.6 & 16 \\
Eclipse of a stated type & 95 & 5 & 19 & 5.2 & 60 \\
Planet against a planet & 78 & 33 & 17 & 5.4 & 0 \\
Solstice or equinox on a day & 54 & 35 & 180 & 2.0 & 26 \\
Sirius date on a day & 41 & 27 & 150 & 2.2 & 41 \\
Acronychal rising on a day & 40 & 19 & 32 & 4.5 & 32 \\
Station against a Normal Star & 31 & 29 & 2 & 8.5 & 0 \\
Planet stations in a sign & 17 & 17 & 51 & 3.8 & 53 \\
\bottomrule
\end{tabular}
\end{table}

Table~\ref{tab:forward} gives the agreement of the reviewed clauses at the published year, by type. The station against a star of the main text agrees in about half of its testable cases. The solstice, equinox and Sirius types agree in nearly all of theirs. These types form the solar-year clock, and their rules approximate the arithmetical scheme (section~S4). Without this clock, the benchmark values stay as in the baseline (section~S7 and Fig.~\ref{fig:ablation}). The two-day lateness of the scheme, set at the published dates, thus has little effect on them.

The authors chose the decisive types from a manual study of the clauses with help from language models. A written rule states its grounds with numbers. A type here is one phenomenon of one body class, such as a distance between two outer planets. A type is physically eligible if it can act as a clock of the year. This leaves out the Moon against a star, the inner planets and the last appearance of a planet. An eligible type is decisive if its clauses agree at the published years in at least 80\% of at least three testable cases.

On the reviewed clauses of all three volumes, the rule gives the decisive types above. Distances between two outer planets, for example, agree in only 3 of 6 testable cases. Lunar eclipses agree in all 3 of theirs. They are also decisive by a fixed choice, because their test uses catalog eclipses and no tolerance. The decisive tier thus holds the clocks that agree at the published years, and the other types help to rank the years.

\begin{table}[H]
\centering
\caption{Agreement at the published year of the 4,010 reviewed clauses from 321 dated records of ADART I to III. A testable clause receives agree or contradict at the published year. Agreement is $100\times$ agree / testable. For the parser clauses alone, from 316 records, 2,169 of 2,313 testable clauses agree (93.8\%).}
\label{tab:forward}
\small
\setlength{\tabcolsep}{4pt}
\begin{tabular}{lrrrrrr}
\toprule
Observation type & Clauses & Testable & Agree & Contradict & Untestable & Agreement (\%) \\
\midrule
Moon or planet against a Normal Star & 2062 & 1330 & 1224 & 106 & 732 & 92.0 \\
Planet in a named sign & 683 & 658 & 643 & 15 & 25 & 97.7 \\
Moon against a planet & 357 & 219 & 203 & 16 & 138 & 92.7 \\
First appearance in a sign & 229 & 228 & 219 & 9 & 1 & 96.1 \\
Last appearance in a sign & 222 & 220 & 207 & 13 & 2 & 94.1 \\
Planet enters a sign & 101 & 101 & 98 & 3 & 0 & 97.0 \\
Eclipse of a stated type & 95 & 5 & 4 & 1 & 90 & 80.0 \\
Planet against a planet & 78 & 32 & 27 & 5 & 46 & 84.4 \\
Solstice or equinox on a day & 54 & 35 & 33 & 2 & 19 & 94.3 \\
Sirius date on a day & 41 & 27 & 27 & 0 & 14 & 100.0 \\
Acronychal rising on a day & 40 & 19 & 18 & 1 & 21 & 94.7 \\
Station against a Normal Star & 31 & 29 & 14 & 15 & 2 & 48.3 \\
Planet stations in a sign & 17 & 17 & 15 & 2 & 0 & 88.2 \\
\addlinespace
All types & 4010 & 2920 & 2732 & 188 & 1090 & 93.6 \\
\bottomrule
\end{tabular}
\end{table}

\subsection*{S6. Scoring and candidate sets (step 3)}

Step 3 reads the verdicts and the tier of each clause, and Table~\ref{tab:roles} lists the quantities that it derives from them. Both depend on settings chosen at the published dates, which the held-out test of section~S8 sets again in each round. Eq.~\ref{eq:weight} applies to a clause with at least one testable and one agreeing year. Other clauses carry zero bits.

\begin{table}[H]
\centering
\caption{The quantities of steps 2 and 3, in the order in which the system uses them for each tablet. Equation numbers refer to the main text.}
\label{tab:roles}
\footnotesize
\setlength{\tabcolsep}{4pt}
\begin{tabular}{@{}p{0.11\linewidth}p{0.08\linewidth}p{0.43\linewidth}p{0.32\linewidth}@{}}
\toprule
Quantity & Symbol & What it is & How it is used \\
\midrule
Verdict & $v_j(y)$ & Agree, contradict or untestable, for clause $j$ in candidate year $y$ & Basis of all later quantities \\
Bits & $u_j$ & $\log_2(C_j/A_j)$, from the numbers of testable and agreeing years of clause $j$ (Eq.~\ref{eq:weight}) & Summed in $I_D$ or $I_S$, shared equally among clauses with identical verdicts \\
Tier & & Decisive if of a decisive type, unrestored, not marked as unobserved, and securely dated or tested over its whole year (section~S5), and supporting otherwise & Only decisive clauses decide which years stay in the strict set \\
Clock & $c$, $K(y)$ & Mars, Jupiter, Saturn, lunar eclipses, or the solar year. $K(y)$ counts the clocks that agree with year $y$ (Eq.~\ref{eq:info}) & Counts independent agreement, apart from bits \\
Summed bits & $I_D(y)$, $I_S(y)$ & Summed bits of the agreeing decisive or supporting clauses (Eq.~\ref{eq:info}) & $I_D$ forms the strict set, and $I_S$ only orders \\
Contradictions & $N_C(y)$ & Number of source statements that contradict the year & Only order the years \\
Strict set & $S$ & Years that no other year beats on clocks and decisive bits (Eq.~\ref{eq:strict}) & The reported set of candidate years \\
Safety set & $S_{\mathrm{safe}}$ & The strict set, joined with the strict sets $S^{(-s)}$ recomputed without each decisive source statement $s$ in turn (Eq.~\ref{eq:safe}) & Holds the years that one misread statement could push out of the strict set. Needed to commit on clocks, and ranks next after the strict set \\
Ranking & $r(y)$ & Strict, then safety, then other years, each by clocks, $I_D$, $I_S$, then fewest contradicting statements (Eq.~\ref{eq:rank}) & Years that tie on all entries form one group \\
First group & $L_1$ & The years tied at the top of the ranking & Tested in the benchmark \\
Decision & $y^\ast$, $E(y^\ast)$ & Commit to the one year $y^\ast$ of the strict set when a decisive clause of an observed lunar eclipse, or two or more clocks with a one-year safety set, confirm it (Eq.~\ref{eq:commit}) & Decline if the record has no decisive clause or the strict set keeps all years, and rank otherwise \\
\bottomrule
\end{tabular}
\end{table}

Clauses with identical verdicts share their bits, as in the main text. If some clauses of such a set are decisive, these share the bits, and the supporting ones add zero bits. Each clause still counts for its own clock in $K$, so Saturn and Mars in Leo count for two clocks. A decisive clause counts for its clock only if it carries bits, that is, if it agrees with fewer years than it can test.

Contradictions are also counted by source statement, so the linked clauses of one statement count once in $N_C$. To form $S^{(-s)}$, the system removes the decisive source statement $s$ with all its linked clauses, and keeps the shared bits of the other clauses. For a record with one decisive source statement, $S^{(-s)}$ holds all 701 years, which widens its safety set.

Tied years form one group, whatever their dates. Years with identical verdicts tie under any score, and a more exact rule or a further clause would separate them. A group can also hold years whose verdicts differ but whose scores tie.

\subsection*{S7. Sensitivity}

Several conventions and rules were set by the authors, some of them at the published dates of ADART I to III. If the results rested on one such choice, a small change to it would move the rates. We therefore change one choice at a time. The sensitivity audit tests 32 fixed configurations, including the baseline, on the same 179 records (Fig.~\ref{fig:ablation}).

For the 120 records with decisive clauses, each configuration reports three values. These are the inclusion of the published year in the first group, the median size of that group, and the inclusion in the strict set. The configurations fall into six kinds, as in Fig.~\ref{fig:ablation}.
\begin{itemize}
\item \textit{Conventions.} A cubit of 2.2 degrees. A shift of ephemeris time by $\Delta T$ at 350 BCE, about 4.2 hours, in all positions. Tolerances narrowed to $2^\circ$ for the Moon and $1.5^\circ$ for a planet, or widened to $4^\circ$ and $3^\circ$. The constant term of Eq.~\ref{eq:britton} moved by $0.5^\circ$ either way. The Moon tested only over the night of a stated day and the night after it, in place of about two and a half days.

\item \textit{Physical evidence.} Removal of Mars, Jupiter, Saturn, lunar eclipses, the solar-year clock, or the Sirius dates alone. A removed clock loses its clauses, together with the distances between the Moon and a removed planet.

\item \textit{Reading rules.} Each run removes one of ten parser rules or rule groups, named in Fig.~\ref{fig:ablation}, together with the clauses that depend on it. The first four give dates or marks. The summary month gives a clause in a month summary the month that ends at the next month heading (section~S2). The restored month context marks a clause that needs a day as restored when its month numeral is restored (section~S5). Its run keeps these clauses but drops the mark. The added month-heading forms read damaged or lowercase month headings, and reading across brackets reads a clause through restored text.

The other six give clause types or names. These are the first appearance, the acronychal rising and the edition's ``Scorpius'' read as the sign Scorpio. The others are the rules from the coding agent (section~S2), a station against a star and a distance between two planets. The coding-agent run removes only parser clauses and keeps the clauses that the search adds. A station against a star and a distance between two planets are supporting types, so their runs can change only the order of the years.

\item \textit{Calendar context.} Removal of all month and day context. Each clause is then tested over its whole year, and clauses that need a day become untestable.

\item \textit{Ordering and strict set.} Contradictions ranked before bits, right after the clocks in Eq.~\ref{eq:rank}. Contradictions counted when forming the strict set, where only its years with the fewest contradicting statements remain.

\item \textit{Scoring.} Joint positions split into separate clauses. Equal counts for all clauses in place of bits, which also removes the tiers. Removal of the tiers alone, so that the bits of all clauses add to one score and no clocks are counted. Equal weights that keep the tiers and clocks. Distances that the rules test only in longitude, such as ``above'' and ``below'', made supporting.
\end{itemize}

Two runs test two approximations in the rules of section~S4, the distances tested only in longitude and the window of the Moon. Making the distances tested only in longitude supporting moves each rate by at most one point. Testing the Moon over two nights in place of two and a half days moves each by at most three.

Two choices of step 3 protect the strict set. When contradictions count in the strict set, or when the tiers are removed, a single misread clause can rule out the published year. The strict set then shrinks to about one year, which is more often wrong. Two other runs rank contradictions before bits or give each clause the same weight within its tier. They give the same rates as the baseline.

Over the 31 settings, the rates change little, and three parts of the method matter most. The tiers protect the strict set, for the reason above. The month context sets the width of the sets and the ranking, because a clause tested over a whole year agrees with many more years. The clocks narrow the years too. Removing one planet more than triples the median strict set. Each planet has a different period, so it cuts the series of years that the others leave.

The conventions matter least, such as the cubit, the zodiac zero point, $\Delta T$ and the tolerances. Each moves a rate by at most three points.

\begin{figure}[H]
\centering
\includegraphics[width=0.8\linewidth]{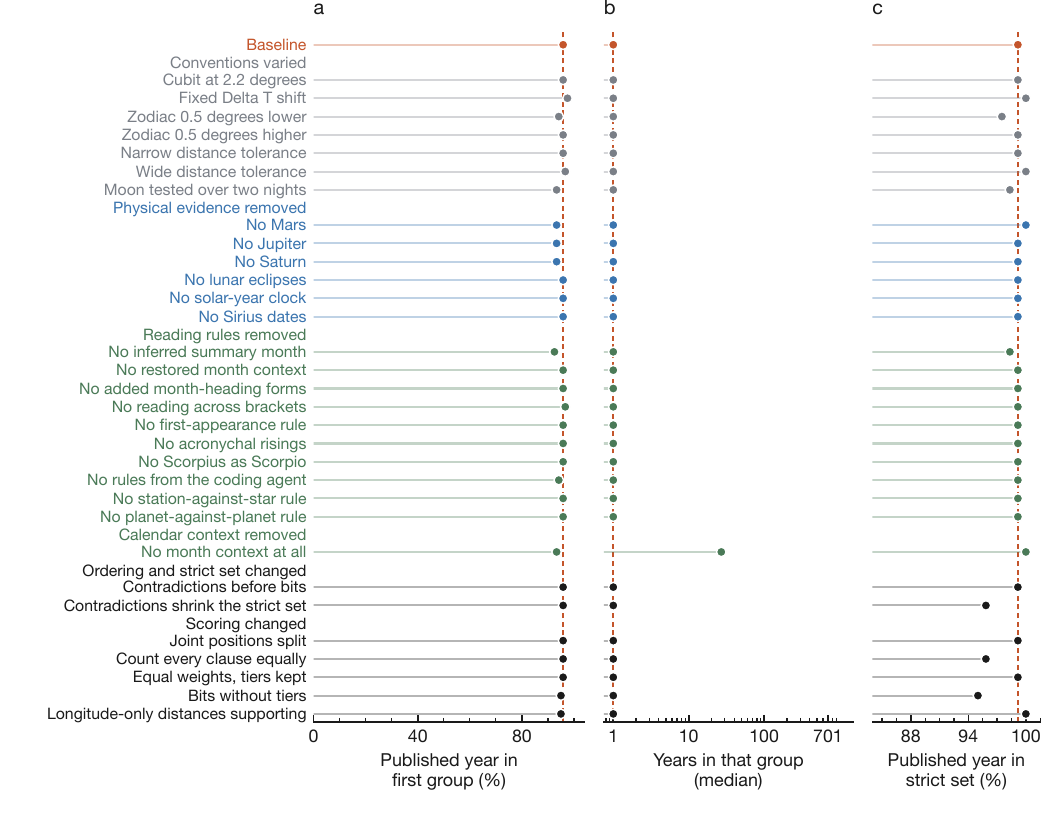}
\caption{Sensitivity of the benchmark for the 120 records with decisive clauses, with the baseline values as dashed orange lines. (a) Published year in the first group, with all tied groups kept. (b) Median number of years in that group, on a logarithmic axis. (c) Published year in the strict set, on an axis that starts above zero. Each row changes one convention, removes one rule or clock, or changes one scoring policy. Group headers mark the six kinds of change, and section~S7 defines each row. The row ``Fixed Delta T shift'' shifts ephemeris time, separately from the three eclipse trials of the baseline.}
\label{fig:ablation}
\end{figure}

\subsection*{S8. Held-out test}

Some settings were chosen at the published dates of ADART I to III, so the benchmark rates could partly come from that choice. A held-out test chose these settings again without the records that tested them. It split ADART I to III by volume and dated the benchmark records of each volume with settings chosen on the other two. The volumes cover different centuries, so each round also tested whether the rules work for another period.

The bits of Eq.~\ref{eq:weight} come from the verdicts, so the settings chosen at the published dates also shape them. A wider tolerance, for example, lets a clause agree with more years and so lowers its bits.

In each round, three kinds of choice were made again from the published years of all dated records in the two training volumes.
\begin{itemize}
\item Decisive types, by the rule of section~S5.

\item Distance tolerances for the Moon and the planets, by the rule of section~S4 applied to the training volumes.

\item Parser rules that entered the code after a check at the published dates (section~S2). A round kept each only if its clauses pass the rule for decisive types.
\end{itemize}
On all three volumes together, these rules give the choices of the main text. Physical eligibility stayed fixed (section~S5). Venus, Mercury, the last appearance of a planet and the Moon against a star stay supporting, and lunar eclipses stay decisive.

Each round computed the verdicts and the bits $u_j$ again. The calendar, the ephemeris, the zodiac relation, the Normal Stars and the eclipse catalogs stayed fixed, because they are independent of the published dates. The windows of the rules, the formulas of step 3, the order of Eq.~\ref{eq:rank} and the benchmark records with their split stayed fixed too. So did the two-day lateness of the solstice scheme, although it was set at the published dates (section~S5). The choices of each round are computed before any test record is scored.

Each round fitted its choices on half to three quarters of the records. The choices thus differ a little between rounds, such as a planet tolerance of $1.5^\circ$ in Round I.

We fixed the pass condition before the test ran. The test uses the pooled rates of the three rounds, for the first group and for the strict set. Each passes if it lies at most five percentage points below the row Rules from all volumes in Table~\ref{tab:heldout}. The margin comes from the size of the benchmark. With 120 records, the 95\% Wilson interval of the first-group rate is 91--98\%, and its lower end lies about five points below the rate. Only the pooled rates carry a threshold, because single volumes hold few records, such as 26 in ADART II.

Table~\ref{tab:heldout} gives, for each test volume, the records with decisive clauses that it holds and their inclusion rates. With settings chosen on the other volumes, both pooled rates equal those of the full benchmark, well within the five points. The results of the main text therefore hold when the settings are chosen without the test records.

A last test used the dated addenda of ADART IV, new to the system, and we registered its rules, input and scores before reading them. The rules then dated each text once, with its parser clauses and from an input built as for the undated entries (section~S9). Texts that join or duplicate a record of ADART I to III are scored apart, and Table~\ref{tab:heldout} gives the rates of the others. The first group holds the published year in 11 of these 12 texts. With so few texts, the 95\% Wilson interval of this rate is wide, 65--99\%, and it includes the benchmark rate of 96\%. In the one miss, the text puts Jupiter in Scorpius where the tools put it just past the boundary, in Sagittarius.

\begin{table}[H]
\centering
\caption{Held-out test by volume. Each of the first three rows dates the benchmark records of one volume with rules set on the other two. Rates are for the records with decisive clauses, as in the main text. All three rounds pools the three test volumes. Rules from all volumes repeats the rates of the main text, where the rules saw all three volumes. The last row gives the registered test on the ADART IV addenda that neither join nor duplicate a record of ADART I to III.}
\label{tab:heldout}
\small
\begin{tabular}{lccccc}
\toprule
Test volume & Records & First group & First three groups & Strict set & Median strict set \\
\midrule
ADART I & 33 & 31 (94\%) & 33 (100\%) & 33 (100\%) & 9 years \\
ADART II & 26 & 25 (96\%) & 26 (100\%) & 26 (100\%) & 49.5 years \\
ADART III & 61 & 59 (97\%) & 60 (98\%) & 60 (98\%) & 17 years \\
All three rounds & 120 & 115 (96\%) & 119 (99\%) & 119 (99\%) & 15.5 years \\
\addlinespace
Rules from all volumes & 120 & 115 (96\%) & 119 (99\%) & 119 (99\%) & 15 years \\
\addlinespace
ADART IV addenda & 12 & 11 (92\%) & 11 (92\%) & 11 (92\%) & 64.5 years \\
\bottomrule
\end{tabular}
\end{table}

\subsection*{S9. ADART IV}

The undated entries of ADART IV show the system at work on texts without a published year. This section describes how they enter the input and what the dating finds. We read the scan of ADART IV \cite{sachs2022adart4} with Mathpix (\url{https://mathpix.com}), and each passage keeps its page and line in the printed volume. Only the undated diaries, on printed pages 95 to 575, enter the input. We split these pages at the printed heading of each entry, which gives 867 entries. Of these, 860 have an English translation.

To keep the dating blind, the input leaves out the printed \textit{Date} and \textit{Comments} sections, the catalog number and the page position. It also leaves out each year, king, ruler or era that the entry names, which a fixed pattern finds. Only the check after dating reads the full translation.

We checked the passages of the decisive clauses against the page images and corrected one passage in which the scan had lost a question mark. Of the 56 ranked entries, 47 hold a single decisive clause. One clock leaves many years, and the first groups of all 56 have a median of 100 years. In 53 of the 56, the decisive clauses lack a month and are tested over the whole year (section~S5). Table~\ref{tab:adart4} marks the other three.

Fig.~\ref{fig:timeline} shows the strict sets of the 56 entries, and Table~\ref{tab:adart4} gives their first three groups. The repository, \url{https://github.com/tingyuansen/Babylonian_Dating}, gives all years of each set and the full ranking of each entry, with its decisive statements.

\begin{figure}[H]
\centering
\includegraphics[width=\linewidth]{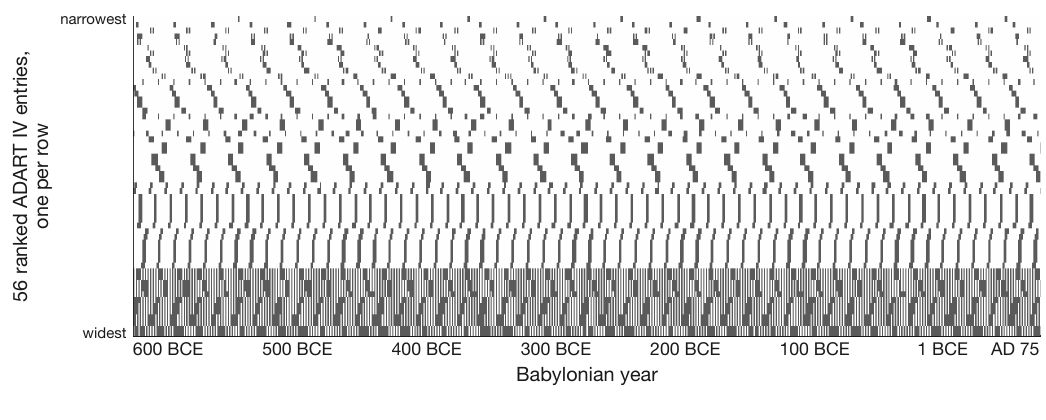}
\caption{Strict candidate sets of the 56 ADART IV entries that keep a decisive clause. Each row is one entry, from the narrowest set at the top to the widest at the bottom, in the order of Table~\ref{tab:adart4}. Dark marks are strict candidates over the 701 years. Marks that repeat at a fixed spacing come from one clock, such as Jupiter, which returns to a sign about every 12 years.}
\label{fig:timeline}
\end{figure}

{\footnotesize
\setlength{\tabcolsep}{4pt}
\setlength{\LTcapwidth}{\linewidth}
\begin{longtable}{@{}p{0.07\linewidth}p{0.05\linewidth}p{0.11\linewidth}p{0.14\linewidth}rrp{0.41\linewidth}@{}}
\caption{Ranked candidate years of the 56 ADART IV entries with a decisive clause, in the order of Fig.~\ref{fig:timeline}. Clocks names the clocks of the decisive clauses, and a dagger marks the three entries with a decisive clause that has a month. $|S|$ and $|L_1|$ count the years in the strict set and in the first group. Groups gives the size of each of the first three groups, or its years when it holds at most ten. Years are in astronomical numbering, so that $-149$ is 150 BCE. The repository gives all years of each group.}\label{tab:adart4}\\
\toprule
Entry & Page & Museum number & Clocks & $|S|$ & $|L_1|$ & Groups \\
\midrule
\endfirsthead
\multicolumn{7}{@{}l}{\textit{Table~\ref{tab:adart4}, continued}}\\
\toprule
Entry & Page & Museum number & Clocks & $|S|$ & $|L_1|$ & Groups \\
\midrule
\endhead
\bottomrule
\endfoot
\bottomrule
\endlastfoot
% Written by scripts/paper/build_adart_iv_table.py; do not edit.
805 & 539 & BM 99671 & Jupiter, Saturn & 21 & 21 & $L_1$: 21 years,\ $L_2$: 75 years,\ $L_3$: 115 years \\
303 & 274 & BM 35863 & Saturn$^\dagger$ & 52 & 52 & $L_1$: 52 years,\ $L_2$: 649 years \\
853 & 567 & MNB 1877 & Mars, Saturn & 54 & 20 & $L_1$: 20 years,\ $L_2$: 34 years,\ $L_3$: $-433$,\,$-228$,\,$9$ \\
264 & 252 & BM 35685 & Mars, Saturn & 56 & 56 & $L_1$: 56 years,\ $L_2$: 28 years,\ $L_3$: 411 years \\
209 & 214 & BM 35294 & Mars, Saturn & 57 & 57 & $L_1$: 57 years,\ $L_2$: 27 years,\ $L_3$: 411 years \\
837 & 558 & Rm 780 & Mars, Saturn & 58 & 58 & $L_1$: 58 years,\ $L_2$: 35 years,\ $L_3$: 378 years \\
77 & 143 & BM 34118 & Mars, Saturn & 58 & 58 & $L_1$: 58 years,\ $L_2$: 38 years,\ $L_3$: 349 years \\
284 & 264 & BM 35765 & Mars, Saturn & 59 & 59 & $L_1$: 59 years,\ $L_2$: 34 years,\ $L_3$: 380 years \\
840 & 560 & Rm 819 & Saturn$^\dagger$ & 59 & 59 & $L_1$: 59 years,\ $L_2$: 642 years \\
602 & 430 & BM 42771 & Mars, Saturn & 61 & 61 & $L_1$: 61 years,\ $L_2$: 39 years,\ $L_3$: 363 years \\
93 & 152 & BM 34364 & Mars, Saturn & 64 & 64 & $L_1$: 64 years,\ $L_2$: 32 years,\ $L_3$: 404 years \\
743 & 503 & BM 47747 & Jupiter$^\dagger$ & 65 & 65 & $L_1$: 65 years,\ $L_2$: 636 years \\
850 & 565 & Rm 1023 & Saturn & 83 & 83 & $L_1$: 83 years,\ $L_2$: 618 years \\
700 & 481 & BM 46066 & Saturn & 83 & 83 & $L_1$: 83 years,\ $L_2$: 618 years \\
288 & 266 & BM 35791 & Saturn & 84 & 84 & $L_1$: 84 years,\ $L_2$: 617 years \\
742 & 503 & BM 47739 & Saturn & 84 & 84 & $L_1$: 84 years,\ $L_2$: 617 years \\
704 & 483 & BM 46087 & Saturn & 88 & 88 & $L_1$: 88 years,\ $L_2$: 613 years \\
651 & 451 & BM 45800 & Jupiter & 89 & 89 & $L_1$: 89 years,\ $L_2$: 612 years \\
37 & 122 & BM 32661 & Saturn & 91 & 91 & $L_1$: 91 years,\ $L_2$: 610 years \\
386 & 317 & BM 36863 & Saturn & 91 & 91 & $L_1$: 91 years,\ $L_2$: 610 years \\
145 & 178 & BM 34893 & Jupiter & 95 & 95 & $L_1$: 95 years,\ $L_2$: 606 years \\
519 & 396 & BM 41957 & Saturn & 96 & 96 & $L_1$: 96 years,\ $L_2$: 605 years \\
371 & 307 & BM 36638 & Saturn & 96 & 96 & $L_1$: 96 years,\ $L_2$: 605 years \\
671 & 462 & BM 45921 & Saturn & 96 & 96 & $L_1$: 96 years,\ $L_2$: 605 years \\
10 & 102 & BM 32223 & Saturn & 100 & 100 & $L_1$: 100 years,\ $L_2$: 601 years \\
427 & 339 & BM 37407 & Saturn & 100 & 100 & $L_1$: 100 years,\ $L_2$: 601 years \\
595 & 428 & BM 42738 & Saturn & 102 & 102 & $L_1$: 102 years,\ $L_2$: 599 years \\
138 & 174 & BM 34862 & Saturn & 102 & 102 & $L_1$: 102 years,\ $L_2$: 599 years \\
325 & 284 & BM 35982 & Saturn & 102 & 102 & $L_1$: 102 years,\ $L_2$: 599 years \\
272 & 257 & BM 35715 & Jupiter & 121 & 121 & $L_1$: 121 years,\ $L_2$: 580 years \\
514 & 394 & BM 41938 & Jupiter & 122 & 122 & $L_1$: 122 years,\ $L_2$: 579 years \\
171 & 193 & BM 35105 & Jupiter & 126 & 126 & $L_1$: 126 years,\ $L_2$: 575 years \\
462 & 362 & BM 41038 & Jupiter & 126 & 126 & $L_1$: 126 years,\ $L_2$: 575 years \\
484 & 374 & BM 41497 & Jupiter & 126 & 126 & $L_1$: 126 years,\ $L_2$: 575 years \\
498 & 386 & BM 41796 & Jupiter & 126 & 126 & $L_1$: 126 years,\ $L_2$: 575 years \\
520 & 397 & BM 41959 & Jupiter & 126 & 121 & $L_1$: 121 years,\ $L_2$: $-465$,\,$-134$,\,$-110$,\,$45$,\,$69$,\ $L_3$: 551 years \\
86 & 149 & BM 34231 & Jupiter & 130 & 130 & $L_1$: 130 years,\ $L_2$: 571 years \\
294 & 269 & BM 35819 & Jupiter & 134 & 134 & $L_1$: 134 years,\ $L_2$: 567 years \\
245 & 237 & BM 35596 & Jupiter & 136 & 136 & $L_1$: 136 years,\ $L_2$: 565 years \\
291 & 268 & BM 35801 & Jupiter & 137 & 137 & $L_1$: 137 years,\ $L_2$: 564 years \\
372 & 307 & BM 36645 & Jupiter & 137 & 137 & $L_1$: 137 years,\ $L_2$: 564 years \\
785 & 529 & BM 77267 & Jupiter & 137 & 137 & $L_1$: 137 years,\ $L_2$: 564 years \\
810 & 541 & BM 99713 & Jupiter & 137 & 137 & $L_1$: 137 years,\ $L_2$: 564 years \\
119 & 164 & BM 34699 & Jupiter & 138 & 138 & $L_1$: 138 years,\ $L_2$: 563 years \\
237 & 233 & BM 35486 & Mars & 419 & 14 & $L_1$: 14 years,\ $L_2$: 47 years,\ $L_3$: 58 years \\
753 & 512 & BM 48151 & Mars & 419 & 58 & $L_1$: 58 years,\ $L_2$: 361 years,\ $L_3$: 38 years \\
184 & 202 & BM 35185 & Mars & 424 & 424 & $L_1$: 424 years,\ $L_2$: 277 years \\
525 & 399 & BM 41976 & Mars & 424 & 424 & $L_1$: 424 years,\ $L_2$: 277 years \\
259 & 249 & BM 35653 & Mars & 424 & 388 & $L_1$: 388 years,\ $L_2$: 36 years,\ $L_3$: 253 years \\
46 & 125 & BM 32765 & Mars & 431 & 431 & $L_1$: 431 years,\ $L_2$: 270 years \\
606 & 432 & BM 42794 & Mars & 436 & 436 & $L_1$: 436 years,\ $L_2$: 265 years \\
747 & 507 & BM 47878 & Mars & 436 & 364 & $L_1$: 364 years,\ $L_2$: 72 years,\ $L_3$: 225 years \\
154 & 182 & BM 34981 & Mars & 448 & 64 & $L_1$: 64 years,\ $L_2$: 384 years,\ $L_3$: 38 years \\
248 & 240 & BM 35599 & Mars & 448 & 448 & $L_1$: 448 years,\ $L_2$: 253 years \\
659 & 455 & BM 45836 & Mars & 462 & 462 & $L_1$: 462 years,\ $L_2$: 239 years \\
684 & 470 & BM 45975 & Mars & 462 & 462 & $L_1$: 462 years,\ $L_2$: 239 years \\

\end{longtable}
}

\FloatBarrier
\subsection*{SI References}
\bibliographystyleS{pnas-new}
\bibliographyS{Bibliography}
\fi

\end{document}